# A Method to Determine Broadband Complex Permittivity of Thin Film Dielectric Materials up to 60 GHz

Liang Wang [1], Guangrui (Maggie) Xia [1, 2], Hongyu Yu [1, 3,4*]

[1] School of Microelectronics, Southern University of Science and Technology, Shenzhen, China
[2] Department of Materials Engineering, The University of British Columbia, Vancouver, BC, Canada
[3] GaN Device Engineering Technology Research Center of Guangdong, Southern University of Science and Technology, Shenzhen, China
[4] The Key Laboratory of the Third Generation Semiconductor, Southern University of Science and Technology, Shenzhen, China

[*] yuhy@sustech.edu.cn

**Abstract:** This paper describes a method to determine the complex permittivity of a thin dielectric film from finite element analysis and microstrip line measurements. Two transmission line equivalent circuit models were used for the cases of an air-filled line and a lossless line, whose distributed elements can be calculated from full wave finite element simulations. With these calculated distributed elements and microstrip line measurements, the complex permittivity was extracted. The technique utilizes a simple way to separate the dielectric loss from the measured total loss and the complex permittivity was extracted using the measured propagation constant. A rational dielectric model was employed to fit the extracted complex permittivity, which ensures causality of the final solution. The best fitting results obtained through this procedure are considered as the final permittivity results, which have shown excellent match to the data sheet values. Moreover, simulations using the fitted permittivity exhibit good agreement with the experimental propagation constant data of microstrip lines up to 60 GHz. The proposed method was demonstrated on polyimide, and it can be applied to other thin film materials.

## 1. Introduction

Current electromagnetic simulation tools can achieve very high accuracy and provide valuable insights on circuit behaviours. This helps circuit designers to build up models for signal analysis and performance predictions. An accurate simulation model requires precise complex permittivities of substrate materials. With the evolution of wireless network towards 5G and beyond, the data transfer speed and bandwidth are required to increase dramatically. When designing high speed electrical circuits, it's critically important to obtain the broadband high frequency dielectric properties instead of using the single point low-frequency permittivity data given by manufacturers.

In the past decades, the extractions of dielectric permittivity have been accomplished by numerous measurement schemes, which mainly fall into three categories: parallel plate capacitors [1], resonators [2, 3] and transmission line (TLine) measurements [4-7]. The parallel plate capacitor method only works at low frequencies and the resonator measurement is reliable at a few discrete resonance frequency points.

The TLine measurements, on the other hand, give repeatable results and are capable of determining frequency-dependent material parameters. Based on TLine measurements, some recent studies relate the imaginary part of the permittivity to the real part using dielectric relaxation models [8,9]. The obtained permittivity is physically consistent, which is described by the Kramers-Kronig relation. However, the algorithm for finding dielectric model parameters is complex. The searching and optimization algorithm normally returns tens or hundreds of solutions, which need to be further selected by numerous simulations and iterations. In this work, we established a more convenient method to address this problem.

The complex permittivity was calculated with the aid of simulations and design equations. Finite element analysis was applied to calculate the lumped element values, which were used to separate the dielectric loss of the substrate from the measured total loss in the microstrip line measurements. At the end, a rational dielectric model was used to fit the calculated raw data, which ensures the causality of the final results. The best-fitting results show good agreement to low frequency data sheet values and were validated by experiments up to 60 GHz.

## 2. Experiment

Polyimide dielectric HD-4110 (from HD Microsystems) was selected as the specimen material, and the thin film microstrip line (TFMSL) was used as the test structure. The TFMSL, as depicted in Fig. 1, consisted of two metal layers and one dielectric layer. In this work, both the signal trace and ground plane were copper films deposited using magnetron sputtering.

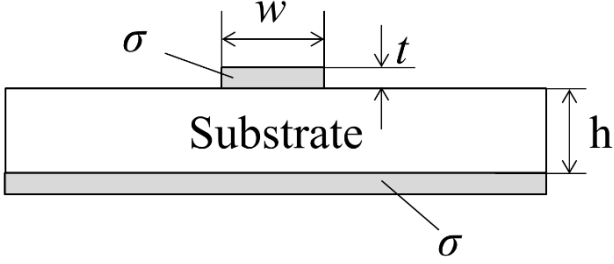

*Fig. 1. Microstrip line cross-section with dimensions and conductivity symbols defined, where w = 38.3, t = 1, h = 14.2 (all in μm) and σ = 5.8×10$^7$ S/m.*

Multiple TFMSLs were fabricated on a single substrate. First, the ground metal layer was deposited directly on the glass carrier and then patterned by a wet etching method. Next, the central polyimide film HD-4110 was spin-coated. In order to make electrical connections from probe pads to the microstrip ground plane, via holes were opened through the central polyimide by a lithography technique. The sample was then cured in a vacuum oven at 375°C. After the curing, a brief reactive ion etch (RIE) treatment was applied on the polyimide surface, and the top metal layer was deposited and patterned. Fig. 2 presents a graph of the fabricated sample consisting a set of TFMSLs with different lengths from 400 to 7000 μm.

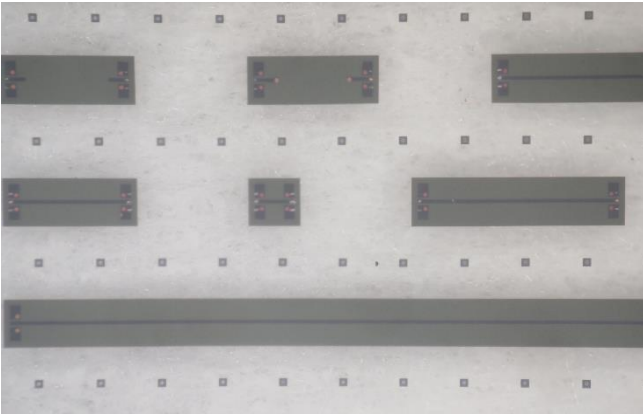

*Fig. 2. Optical image of the fabricated TFMSLs for the extraction of polyimide permittivity at x10 magnification.*

All the lines, in addition to connecting with the same Ground-Signal-Ground (GSG) probe pads, were designed to have the same line width of 38.3 μm. A linear taper transition was made from the central probe pad to the microstrips. The thickness of polyimide was measured to be 14.2 ± 0.1 μm across the whole surface area. After the fabrication, the surface roughness of both polyimide and the top metal trace were inspected Wyko optical profilometer. The calculated root-mean-square roughness for both polyimide and metal surface was below 20 nm, which was low enough that can be neglected in the following analysis.

The microwave measurements on the TFMSLs were carried out using a vector network analyser (VNA) AGT E8361C from Agilent Technologies. A total number of 401 frequency points was measured over the frequency range from 100 MHz to 60 GHz. A pair of GSG probes with a pitch of 150 μm were connected to the VNA through coaxial cables. After an initial probe tip calibration, the probes were directly placed on the TFMSL contact pads to start measurements. The temperature of laboratory during measurement was 20 °C.

Generally, propagation constants can be determined from the scattering parameter measurements of transmission lines using multiline method [10]. If two measured microstrip lines have the same line width and dielectric thickness, a larger difference in their line length can minimize the error in calculating the propagation constant. However, it is difficult to control the dimension uniformity for an extremely long line. Any variation in the dimension will deteriorate the accuracy of the extraction. In this work, we used only two TFMSLs with the lengths of 400 and 7000 μm respectively.

### 3. Distributed Element Model

Below, we will discuss the first step of our method, where simple equivalent circuit models were used to calculate the important distributed elements such as the resistance per unit length $R$, the inductance per unit length $L$, the external inductance per unit length $L_{\text{ext}}$ and the air-filled capacitance per unit length $C_a$, which are illustrated in Fig. 3.

The complex permittivity $\varepsilon$ of a dielectric material is defined as

$$\varepsilon = \varepsilon_0 \varepsilon_r (1 - j\tan\delta) \quad (1)$$

where $\varepsilon_0$ is the free space permittivity, $\varepsilon_r$ is the relative permittivity and $\tan\delta$ is the loss tangent.

When the TFMSL operates in the quasi-TEM region, a well-known distributed equivalent circuit model can describe its transmission behaviour, as shown in Fig. 3*a* [11]. The TLine characteristic impedance $Z$ and the propagation constant $r$ can be written in terms of the distributed elements:

$$Z = \sqrt{\frac{R + j\omega L}{G + j\omega C}} \quad (2)$$

$$r = \sqrt{(R + j\omega L)(G + j\omega C)} \quad (3)$$

where $\omega$ denotes the angular frequency.

The resistance per unit length $R$ and inductance per unit length $L$ are independent of the central dielectric property [12]. Thus, in case of an air-filled line, the equivalent circuit model can be simplified by removing the shunt conductance $G$ while retaining the series impedance elements $L$ and $R$, as shown in Fig. 3*b*. It should be noticed that the capacitance per unit length $C$ is replaced by the air-filled capacitance per unit length $C_a$.

It is been proved that the total inductance $L$ is a sum of its internal and external part [13]. The internal inductance $L_{\text{int}}$, which stems from the magnetic field inside the conductors, can be further excluded in the distributed model for an air-filled lossless line as shown in Fig. 3*c*.

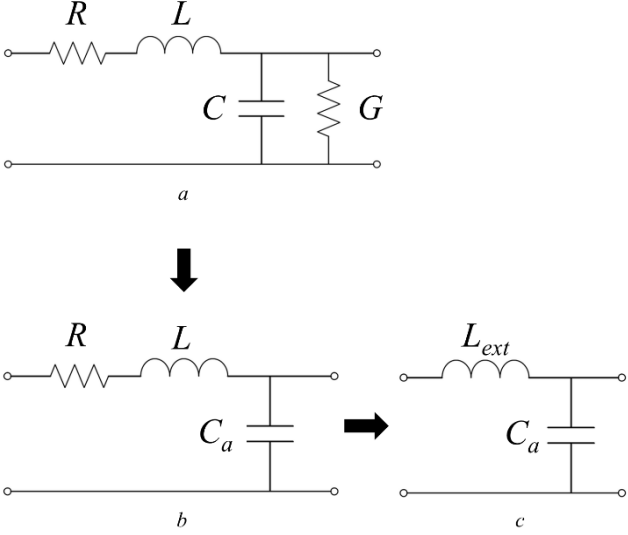

*Fig. 3. Distributed equivalent circuit models of the microstrip lines for the following cases: (a) a dielectric-filled line, (b) an air-filled line, and (c) an air-filled lossless line.*

Because the expressions in (2) and (3) are also valid for the equivalent circuit models in Figs. 3*b* and *c*, these two expressions can be used to determine the distributed element values once we know the corresponding Z and $r$. With the aid of a full wave simulation, the frequency-dependent Z and $r$ of a TLine can be calculated. Then by combining (2) and (3), we can determine the value of the distributed elements such as $L$, $R$, $L_{ext}$ and $C_a$.

In this work, Ansoft HFSS was employed to perform the simulations. Since the surface roughness of the TFMSL was within tens of nanometres, its influence on transmission was too small to be incorporated in the simulation. First, two air-filled microstrips were built with the same geometry as shown in Fig. 1. In the first structure, the metal resistivity was defined as zero, and therefore the equivalent circuit is in Fig. 3*c*. In the second structure, the Cu metal conductivity was used where $\sigma = 5.8 \times 10^7$ S/m, and therefore the equivalent circuit is in Fig. 3*b*. Then, Z and $r$ of each structure were calculated by Ansoft HFSS. After that, $L_{ext}$ and $C_a$ were solved from Z and $r$ of the first structure, and $R$ and $L$ were solved from Z and $r$ of the second structure. Fig. 4 plots the calculated $L$ and $L_{ext}$. Fig. 5 shows the results of $R$ and $C_a$.

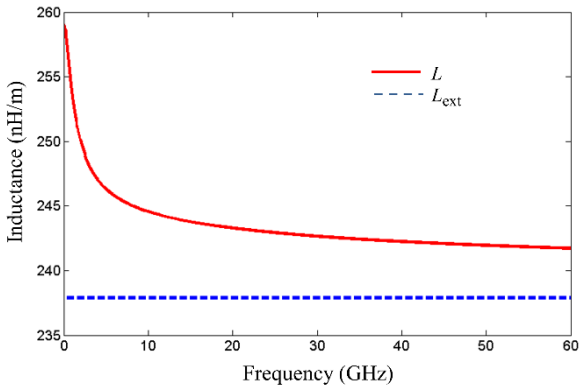

*Fig. 4. Calculated L and $L_{ext}$ for TFMSL.*

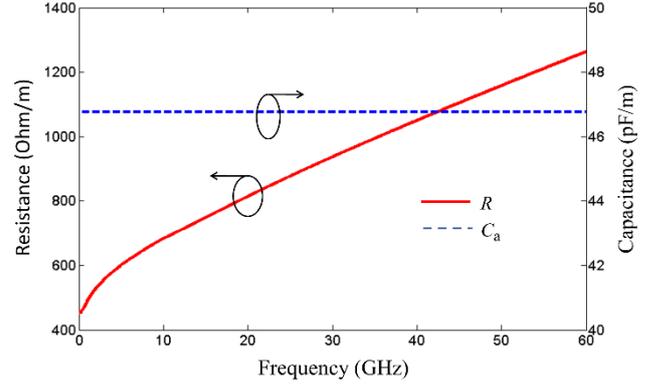

*Fig. 5. Calculated R and $C_a$ for TFMSL.*

## 4. Dielectric Permittivity Extraction

The next step was to extract the complex permittivity of the dielectric substrate. The propagation constant $r$ of the fabricated TFMSL was calculated using two lines with different lengths. For each line, the measured two port scattering matrix was converted to a form of cascading matrix $M$, which is consider as a product of three parts:

$$M_1 = X T_1 Y \quad (4)$$

$$M_2 = X T_2 Y \quad (5)$$

where $X$ is the input matrix including the input transition from probe tip to microstrip line, $T$ represents the transmission line matrix, and $Y$ is the output matrix representing the output transition from microstrip to probe tip. The matrix $X$, $T$ and $Y$ are all in cascading form. If multiplying the matrix $M_1$ by the inverse matrix $M_2$, the combination is an eigenvalue equation:

$$M_1(M_2)^{-1} = X T_1 (T_2)^{-1} (X)^{-1} \quad (6)$$

Using the fact that the sum of the diagonal elements does not change under the similar transformation in matrix calculation, the propagation constant can be derived from the relation

$$\text{tr}(M_1(M_2)^{-1}) = \text{tr}(T_1(T_2)^{-1}) = 2\cosh r \Delta l \quad (7)$$

where $\Delta l$ is the length difference between two transmission line.

Once the propagation constant $r$ is obtained, we can separate it into real part and imaginary part. The real part $\alpha$ is called attenuation constant, which reflects the loss of propagation; the imaginary part $\beta$ is called phase constant, which represents the speed of transmission.

When an electromagnetic wave propagates in a microstrip transmission line, most of the energy contains in the dielectric substrate, and a portion of the energy exists in the ambient air. To account for this effect, the notions of complex effective permittivity $\varepsilon_{eff}$ and effective permeability $\mu_{eff}$ are introduced. Therefore, the propagation constant can be written as a function of $\varepsilon_{eff}$ and $\mu_{eff}$ as given in [14]

$$r = \alpha + j\beta = j\frac{\omega}{c}\sqrt{\mu_{eff}\varepsilon_{eff}} \quad (8)$$

where $c$ is the speed of light.

Combining equation (8) and (3), we can deduce the expression of $\varepsilon_{\text{eff}}$ and $\mu_{\text{eff}}$ in terms of the lumped elements

$$\mu_{\text{eff}} = \frac{R + j\omega L}{j\omega L_{\text{ext}}} \quad (9)$$

$$\varepsilon_{\text{eff}} = \frac{G + j\omega C}{j\omega C_a} = \varepsilon'_{\text{eff}} - j\varepsilon''_{\text{eff}} \quad (10)$$

where $\varepsilon'_{\text{eff}}$ and $\varepsilon''_{\text{eff}}$ are the real and imaginary part of $\varepsilon_{\text{eff}}$. Then combining equation (8) to (10) yields

$$\beta^2 - \alpha^2 + j2\alpha\beta = \frac{\omega^2}{c^2}\left(1 + \frac{L_{\text{int}}}{L_{\text{ext}}}\right)\varepsilon'_{\text{eff}} + j\frac{\omega^2}{c^2}\left(\frac{L}{L_{\text{ext}}}\frac{G}{\omega C_a} + \frac{\varepsilon'_{\text{eff}}R}{\omega L_{\text{ext}}}\right) \quad (11)$$

In order to extract the complex permittivity, the first step is to find its relative permittivity $\varepsilon_r$. By taking the real part of (11) and reorganize it, $\varepsilon'_{\text{eff}}$ can be written as

$$\varepsilon'_{\text{eff}} = \frac{c^2}{\omega^2}\frac{\beta^2 - \alpha^2}{\left(1 + \frac{L_{\text{int}}}{L_{\text{ext}}}\right)} \quad (12)$$

Since both the $L_{\text{int}}$ and $L_{\text{ext}}$ are already known from the full-wave simulations, $\varepsilon'_{\text{eff}}$ then can be determined from (12).

It is obvious that there is one-to-one correspondence between $\varepsilon'_{\text{eff}}$ and the relative permittivity $\varepsilon_r$. If the microship linewidth $w$ is larger than its dielectric thickness $h$, which is true for our TFMSL, $\varepsilon_r$ can be found through the relation given in [8]

$$\varepsilon'_{\text{eff}} = \frac{\varepsilon_r + 1}{2} + \frac{\varepsilon_r - 1}{2}\left(1 + \frac{12h}{w}\right)^{-\frac{1}{2}} - 0.217(\varepsilon_r - 1)\frac{t}{\sqrt{wh}} \quad (13)$$

The second step is to calculate the dielectric loss tangent $tan\delta$. The imaginary part of (11) can be written as

$$2\alpha\beta = \frac{\omega^2}{c^2}\left(\frac{L}{L_{\text{ext}}}\frac{G}{\omega C_a} + \frac{\varepsilon'_{\text{eff}}R}{\omega L_{\text{ext}}}\right) \quad (14)$$

When the microwave energy propagates in a microstrip line, the attenuation caused by dielectric loss is modelled by the shunt conductance $G$. On the other hand, the energy stored in dielectric material results in shunt capacitance $C$, which can be further expressed by substrate relative permittivity and air-filled capacitance $C_a$. According to [11, p.155], the relation between conductance $G$ and $C_a$ is given as

$$\frac{G}{\omega C_a} = \frac{(\varepsilon'_{\text{eff}} - 1)\varepsilon_r}{\varepsilon_r - 1}tan\delta \quad (15)$$

By substitute (15) into (14), one has

$$\alpha = \frac{\omega^2}{2\beta c^2}\frac{L}{L_{\text{ext}}}\frac{(\varepsilon'_{\text{eff}} - 1)\varepsilon_r}{\varepsilon_r - 1}tan\delta + \frac{\omega^2}{2\beta c^2}\frac{\varepsilon'_{\text{eff}}R}{\omega L_{\text{ext}}} \quad (16)$$

The first term on the right hand of the (16) is related to the dielectric loss, and the second term represents the conductor loss. It is now possible to calculate the value of the loss tangent using (16), since all the other parameters in (16) have already been obtained throughout the above procedure.

The solid lines in Fig. 6 and 7 show the extracted broadband substrate $\varepsilon_r$ and $tan\delta$. The dashed lines in both figures are the best-fitting curves using Equation (17), which will be further discussed in the next section.

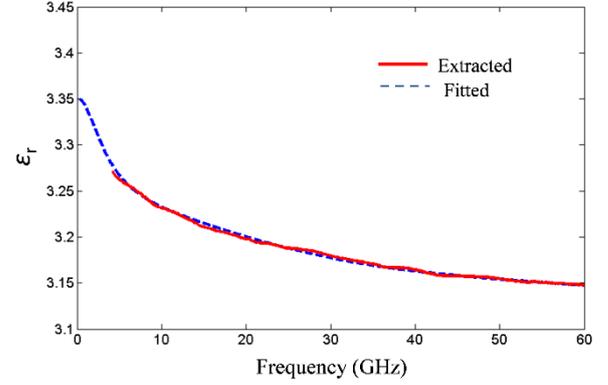

*Fig. 6. The extracted $\varepsilon_r$ of polyimide HD-4110 and the best-fitting results with Equation (17).*

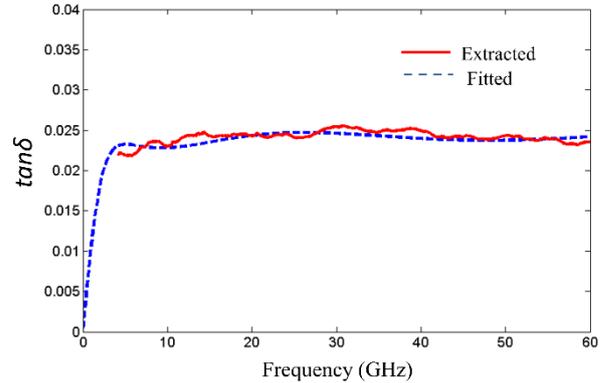

*Fig. 7. The extracted $tan\delta$ of polyimide HD-4110 and the best-fitting results with Equation (17).*

As shown by the solid lines, the extracted data below 4.2 GHz were omitted, because the strong slow wave effect is dominant for this frequency range. The slow wave effect, which is caused by small ratio of the metal thickness to the skin depth, makes permittivity extraction has large variation [7]. At about 4.2 GHz, the metal thickness is equal to the skin depth. Therefore, we can assume that this method is accurate beyond this threshold frequency.

## 5. Dielectric Model Fitting and Validations

The final results for the dielectric permittivity must also satisfy causality. A rational permittivity model, which consists of a discrete sum of dipole moments, is given as [9]

$$\varepsilon(j2\pi f) = \varepsilon_\infty + \sum_{n=1}^{\infty}\left(\frac{c_n}{j2\pi f - a_n}\right) \quad (17)$$

where $f$ is the frequency, $\varepsilon_\infty$ denotes the infinite frequency limit, $a_n$ is the poles and $c_n$ is the residues. For the sake of stability, the real part of $a_n$ must be negative.

This dielectric model was employed to fit the extracted permittivity using Matlab. The number of poles in the dielectric model should be kept small while it still could generate good fittings. For the concerned bandwidth in this work, four-pole equation is sufficient to express the dielectric function. It should be mentioned that when the fitting is required for a broader bandwidth, the number of dipoles may need to be increased accordingly. The result of the fitting is plotted as dashed lines in Fig. 6 and 7. These best-fitting curves are based on Kramers-Kronig relation and were calibrated by the experimental results. Therefore, we consider these results are our final permittivity results.

It can be seen that the extracted permittivity can be well approximated by the model. Moreover, the best-fitting model can extrapolate the permittivity value below 4.2 GHz. At the lowest frequency of 1 MHz, the model predicted a relative permittivity of 3.35, which is very close to 3.36 provided by the manufacturer at the same frequency. The same consistency was also found for the loss tangent. The best-fitting data predicted that below 100 MHz the loss tangent of HD-4110 should be less than $1.1\times10^{-3}$, while that specified by the manufacturer is $1\times10^{-3}$ at 1 MHz.

As the good agreement with data sheet values was in the low frequency regime, we needed another validation at the high frequency regime. To do that, a new pair of TFMSLs were produced using the same thin film technology. The propagation constant $r$ was measured up to 60 GHz and compared with the simulated results using the permittivity data we just obtained for polyimide HD-4110.

The new TFMSLs have a different thickness of 7.1 μm and a different line width of 25.5 μm. An HFSS simulation of the new TFMSL structure using the permittivity data was performed to calculate $r$. The simulated $r$ was compared to the measured value. Fig. 8 shows that a good agreement is achieved between the measurement and HFSS simulation results across the whole bandwidth up to 60 GHz.

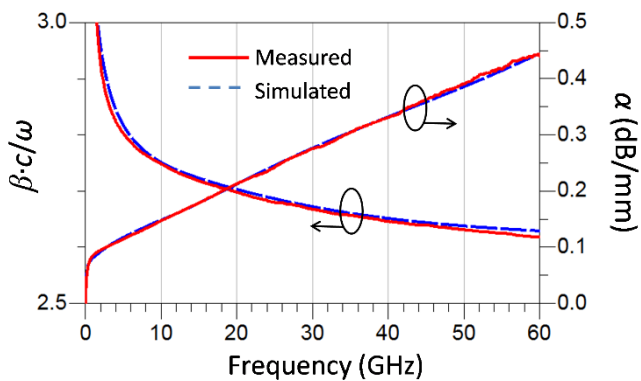

*Fig. 8. Comparison of the measured and simulated propagation constant for TFMSL with dimension of w = 25.5, t = 1, h = 7.1 (all in μm) and σ = 5.8×10⁷ S/m.*

## 6. Analysis and Discussion

Although the above technique shows promising results, some factors can influence the accuracy of the method. The best-fitting curve relies on the accuracy of the raw permittivity data, which were originally calculated with the aid of HFSS simulations. Therefore, errors in the simulation structures and settings, such as surface roughness, variations of dimensions, uncertainty in metal conductivity, may cause deviations of results from the true values. However, these factors can be minimized by using the average value of measurements at multiple sites along the TLines.

## 7. Conclusion

In this work, a simple procedure was proposed and validated to determine physically consistent complex permittivity of a thin film dielectric material from very low frequency up to 60 GHz. Knowing the geometry of the microstrip lines, distributed element values in the equivalent circuit model were obtained by merely two simulations. Based on the obtained distributed elements and the measured propagation constant, the dielectric loss was separated from the measured total loss. The extracted complex permittivity was well fitted by a rational dielectric model. The best-fitting curve satisfies the causality constraints and was considered as final result, whose values match to the available data sheet values. A good agreement was also demonstrated between the experimental and the simulated results up to 60 GHz.

## 8. Acknowledgment

This work was supported in part by Science, Technology and Innovation Commission of Shenzhen under the Contract of JCYJ20160226192639004, JCYJ2017-0412153356899 and was supported in part by Department of Science and Technology of Guangdong Province under the Contract of 2017A050506002, 2019B010128001